\documentclass[journal]{IEEEtran}
\ifCLASSINFOpdf
\else
\fi
\usepackage{graphicx}
\begin{document}
%
\title{Deep Modulation Embedding}
%
%
%

\author{Amin Abbasloo, and
        Alan Salari
\thanks{Amin Abbasloo and Alan Salari are with DRONEASURE.

E-mail: \{abbasloo,salari\}@droneasure.com}
\thanks{Manuscript received on February, 2019; revised on April, 2019.}}

%
%

\markboth{Journal of X,~Vol.~Xx, No.~Xy, April~2019}%
{Shell \MakeLowercase{\textit{et al.}}: Bare Demo of IEEEtran.cls for IEEE Journals}
%



\maketitle

\begin{abstract}
Deep neural network has recently shown very promising applications in different research directions and attracted the industry attention as well. Although the idea was introduced in the past but just recently the main limitation of using this class of algorithms is solved by enabling parallel computing on GPU hardware. Opening the possibility of hardware prototyping with proven superiority of this class of algorithm, trigger several research directions in communication system too. Among them cognitive radio, modulation recognition, learning based receiver and transceiver are already given very interesting result in simulation and real experimental evaluation implemented on software defined radio. Specifically, modulation recognition is mostly approached as a classification problem which is a supervised learning framework. But it is here addressed as an unsupervised problem with introducing new features for training, a new loss function and investigating the robustness of the pipeline against several mismatch conditions.        
\end{abstract}


%
\IEEEpeerreviewmaketitle

\section{Introduction}
%
%
%
%
\IEEEPARstart{N}{eural} network was an active research area in the past and reached a point where no more investigation could be done because of its needs of computational resources for the training. Just recently, a few hardware companies released affordable GPU hardware that enables doing heavy computation in parallel. Consequently, API's and compilers were developed and made public for handling the required computation for their training on GPU. This opened an opportunity for investigating the potentials of neural network even with more layers (deep network). Within few years, many different applications, new architectures and numerical techniques were suggested to address a wide range of tasks and fixing numerical issues of training the networks. Deep neural networks (DNN) showed incredible result outperforming the state-of-art algorithms and even pushed machine leaning techniques to a new level as well introduced new use cases that were possible only by human \cite{IEEEhowto:LeCun}. Computer vision and graphics, speech and data science are the success stories of this newborn class of algorithms. 

Electrical engineers have tested successfully the machine learning techniques for designing and analyzing the communication systems and signals. By re-introducing DNN, it was quite natural for communication system community to define projects for exploring the possibilities that this new paradigm can offer and even see the industrial applications \cite{IEEEhowto:O'Shea0}. Moreover, NVIDIA released a new hardware family called 'Jetson' that enables hardware prototyping suited for communication applications. The testbed is basically made of a software defined radio (SDR), Jetson GPU server and FPGA. SDR enables receiving and transmitting signals over channels and FPGA does the required computation real-time and fast. Jetson GPU server can hold the already trained network using a PC or laptop and later be stacked to the testbed hardware by a carrier for testing phase and real-world applications.

Among many research projects that have been launched, a few like cognitive radio, modulation recognition and learning based receiver, transceiver got the most attention although soon or later new applications will be out. The main difference between classical and DNN based method is, classical method are designed based on the expert knowledge but DNN based ones are basically data driven. The network learns during training the data space and can react accordingly during testing phase when it is fed by new/unseen data. Employing DNN in communication system can be done by improving the state-of-art design made by expert knowledge by replacing some blocks and modules by DNN. Moreover, one can also design the whole system based on DNN ignoring already established designs. Designing new DNN based receivers (keeping transmitter as a current state-of-art design) or a totally DNN based transceiver that both transmitter and receiver are replaced by DNN and trained over the channel are done/reported improving the capacity and reduced the bit error rate.

As the focus of this work is on modulation recognition, literature review mainly contains the already established and reported directions for cognitive radio and modulation recognition. Cognitive radio tries to monitor and manage the spectrum between users so it is required to detect the unused frequencies and identify the users (including the modulations). Detection is basically based on signal energy and modulation recognition is formulated by some expert knowledge and hand-crafted features like cyclostationary signatures in classical methods \cite{IEEEhowto:Dobre}. A few publications have addressed the modulation recognition as a classification problem. It is a supervised leaning that DNN learns the data space by knowing the ground truth as labels (each label shows a specific modulation). Using DNN for classification is hugely tested in image processing community and it also proved its performance in modulation recognition. Although there might be ambiguity for some modulations but a few extra steps for them can solve the issue \cite{IEEEhowto:O'Shea1, IEEEhowto:O'Shea2}. Formulating the problem as an unsupervised learning is a flexible platform to deal with unlabelled data and provides a signature/feature for signals which can be used later for different tasks like classification, clustering and detection \cite{IEEEhowto:Usama, IEEEhowto:O'Shea3}. This perspective is not properly studied before and there are several open questions regarding the architecture, features and the robustness of the unsupervised techniques for modulation recognition. These give motivation for doing a systematic investigation along this line of work.    

Section II explains the DNN architecture used for investigation and provides more details about loss function based on Markov modelling and new features for the training. Section III presents the training setup and discuss the experiments that have been done in order to see the effects of free parameters and robustness of the algorithm in mismatch conditions. A very cheap and affordable testbed has been introduced and implemented by HackRF One and GNURADIO.

\section{Network Architecture}
Variational autoencoder is used as the network architecture for signal embedding. This is usually a natural choice for dimensionality reduction and has been used for denoising too \cite{IEEEhowto:Kingma}. The network is supposed to embed the input data with dimension $N$ in a new space with $M$ such that $N>M$, here $M=1$ in Fig. 1. The loss function can be defined in order to include more terms like Markov modelling \cite{IEEEhowto:Hernandez}. During the optimization, the difference between the shifted signal and the decoder output with different lags is minimized, it has been inspired by Markov modelling (Appendix A). Features are I/Q components themselves, differences between I/Q components and their shifted versions (because the signal trajectory also shows some characteristics of the signal) and correlation with different lags.

\begin{figure}[ht]
  \centering
    \includegraphics[width=8.0cm,height=3.0cm]{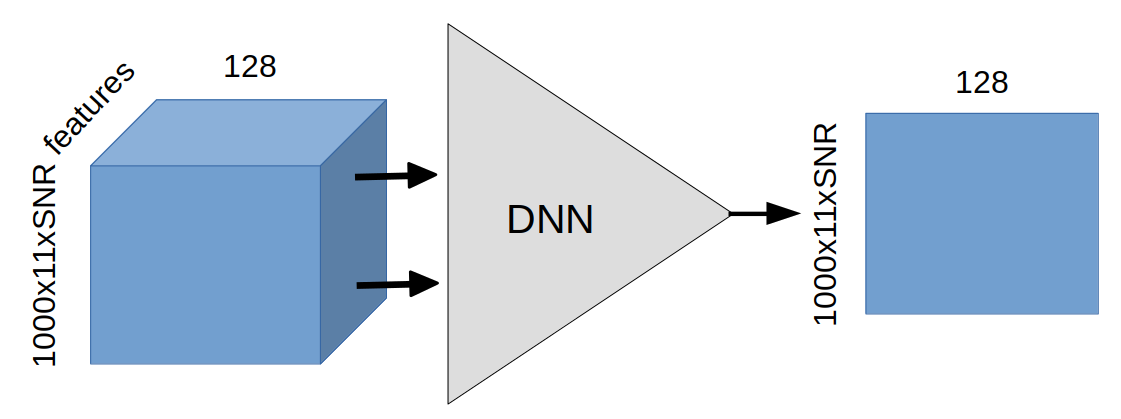}
    \caption{Network's input and output dimension.}
\end{figure}

\section{Result and Evaluation}
Dataset is obtained from \cite{IEEEhowto:deepsig} and was generated using GNURADIO. Although, DNN usually needs GPU for training but the network architecture used here can be/was trained on CPU. The dimension of hidden layer was $256$ with depth equals to $3$, $10^{-3}$ learning decay and $0.2$ dropout. Since the network is pretty small and trains very fast even on CPU, it actually gives the opportunity to use it for primary evaluation with little budget. Its implementation is available here \texttt{https://github.com/abbasloo/dnnSDR}.

\subsection{Offline}
The network is trained on eleven different modulations with different SNR's each having 1000 measurement with 125 samples. For visualizing the result, I/Q trajectory is color coded with embedded space to show the signal characteristics and with a two-dimensional histogram of difference between embedded space and its first lag shifted versus embedded signal itself as a simple signature.
After training, the network has been tested on unseen data for different SNR's and modulations. Fig. 3 shows the result for a few seen SNR's, each row corresponds to a specific SNR. Noted that, for example from Fig. 3, ('WBFM', '10') means the modulation is 'WBFM' with SNR equals to 10 db. By decreasing the SNR, the signal characteristics changes gradually, Fig. 3 from top to bottom. Lag differences  with areequals to eight lag only features used for training for result in Fig. 3. Since lag differences and correlations are introduced as new features here, therefore the affect of increasing lag and including extra features are investigated too. In Fig. 4, first row shows the normalized distance between embedded spaces with lag equals to eight and sixteen. The second row is the distance between lag differences plus correlations with lag equals to eight and just lag differences with lag equals to sixteen as features (which both give more or less the same features dimension). Increasing lags and including correlation have impacts on all modulation and change the two-dimensional histograms too. This shows that changing parameters and including or excluding some features, might help to embed visually similar signals but with different characteristics in a new space that is discriminating enough. Third row shows the distance if 'WBFM' is included or excluded during the training. The network is able to create the signal characteristics even though it has not seen the modulation during the training phase. But it has a strong impact on other modulations as the concentration of warm color is increased for other modulations in the graphs. As it can be seen from Fig. 4 the samples close to I/Q space center are not very robust for different conditions but those near the edge are very strong and carry the modulation characteristics. The forth row presents the distance if one specific SNR (6 db) is included or excluded during training (training data includes SNR's up to 6 db, Fig. 3). Changing the conditions in general influences the modulations differently even for those with similar I/Q trajectories like 'AM-SSB' and 'QAM16'.

\subsection{Online}
A GNURADIO data flow is set up to receive the data from a SDR and acts as a publish/subscribe server (Fig. 5). The SDR hardware is a HackRf One which is very cheap and affordable as well provides the possibility for developing a reasonable/low budget prototype for DNN based project in communication systems specially for student research projects. The recorded signal has been fed to the trained network. Fig2 shows the signal trajectory color coded with embedded space and two dimensional histogram as a simple signature. The station is a local FM radio station and its modulation is 'WBFM'. The histogram shows a flip along y-axis comparing to Fig. 3 for 'WBFM'. It is because of the rotation ambiguity of the embedded space.

\begin{figure}[ht]
  \centering
    \includegraphics[width=4.0cm,height=2.0cm]{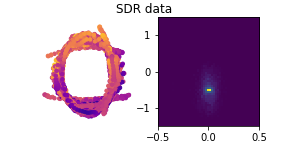}
    \caption{local FM radio signal with 102.4 MHz frequency in Bonn region. Left image shows signal trajectory color coded with embedded space and right is its two dimensional histogram.}
\end{figure}

\begin{figure*}
  \centering
    \includegraphics[width=18.0cm,height=1.8cm]{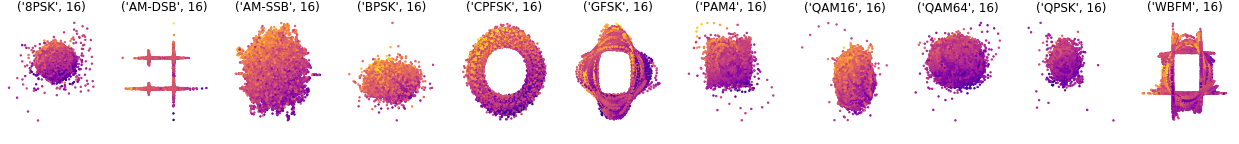}
    \includegraphics[width=18.0cm,height=1.8cm]{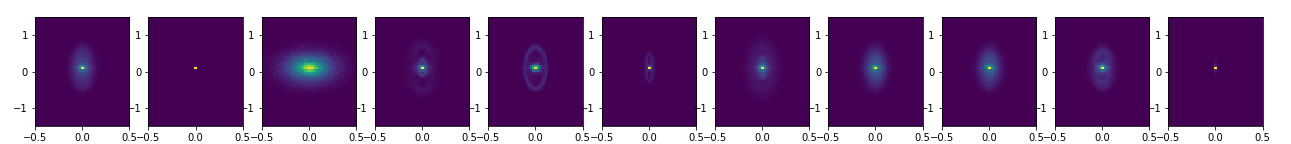}
    \includegraphics[width=18.0cm,height=1.8cm]{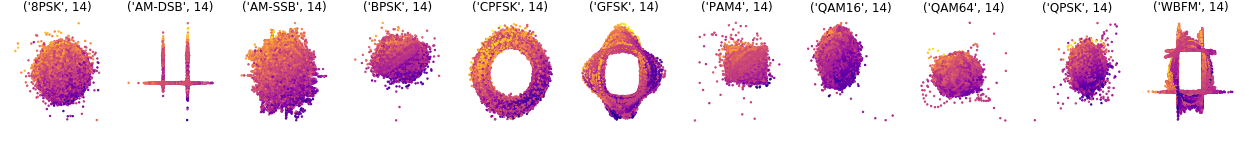}
    \includegraphics[width=18.0cm,height=1.8cm]{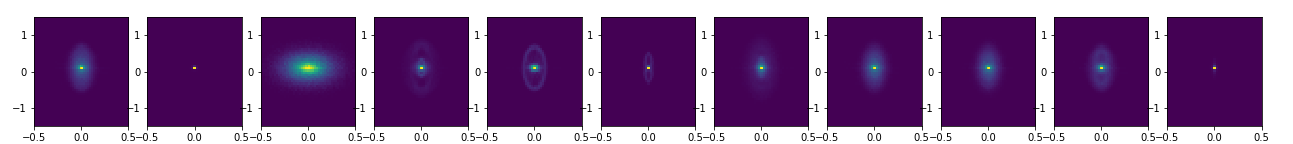}  
    \includegraphics[width=18.0cm,height=1.8cm]{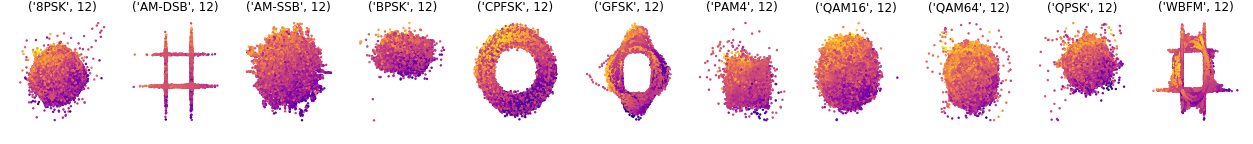}
    \includegraphics[width=18.0cm,height=1.8cm]{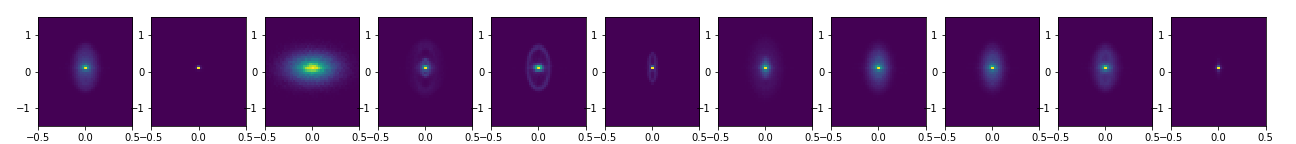}    
    \includegraphics[width=18.0cm,height=1.8cm]{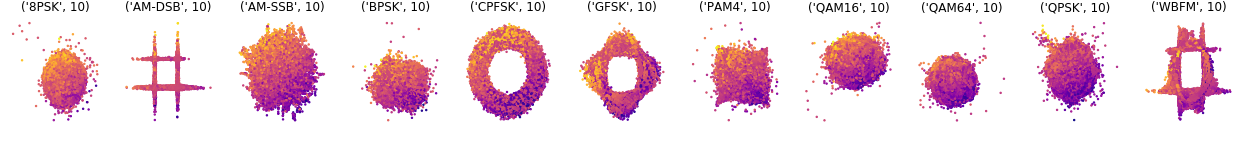}
    \includegraphics[width=18.0cm,height=1.8cm]{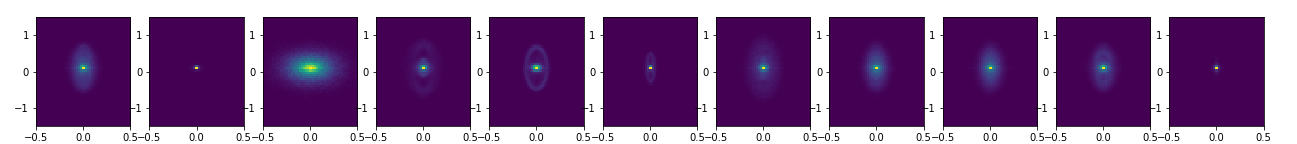} 
    \includegraphics[width=18.0cm,height=1.8cm]{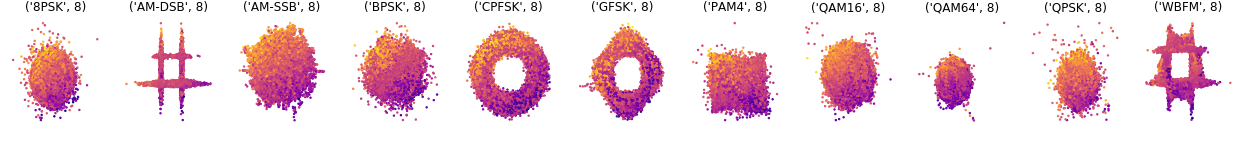}
    \includegraphics[width=18.0cm,height=1.8cm]{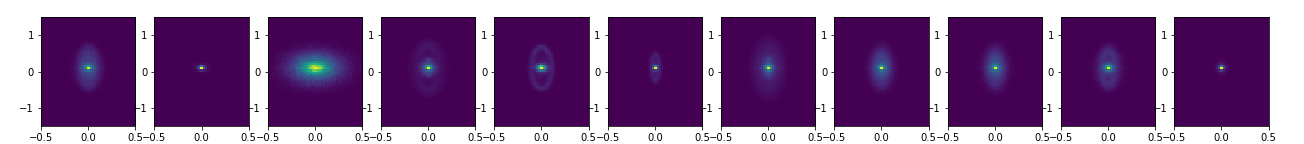}   
    \caption{Five double-rows (one row shows the signal trajectory color coded by embedded space and the other is its two dimensional histogram of difference between embedded signal and its first lag shift versus embedded signal itself) show signal trajectory for eleven different modulations (each row has eleven modulations and corresponds to a specific SNR). By decreasing the SNR, the two dimensional histogram, as a simple signal signature, slightly changes. }
\end{figure*}

\begin{figure*}
  \centering
    \includegraphics[width=18.0cm,height=1.8cm]{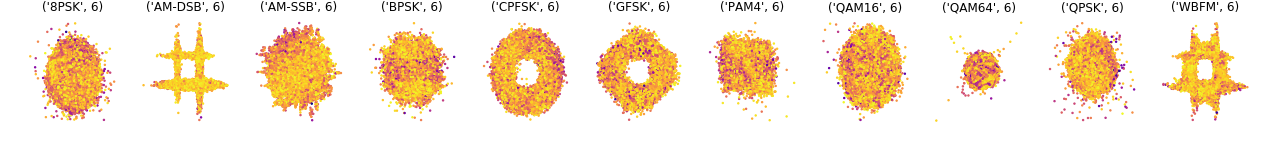} 
    \includegraphics[width=18.0cm,height=1.8cm]{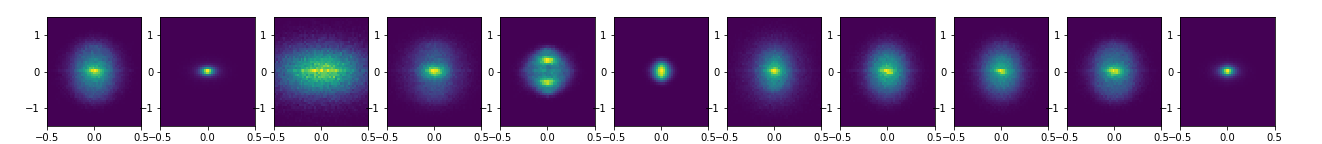} 
    \includegraphics[width=18.0cm,height=1.8cm]{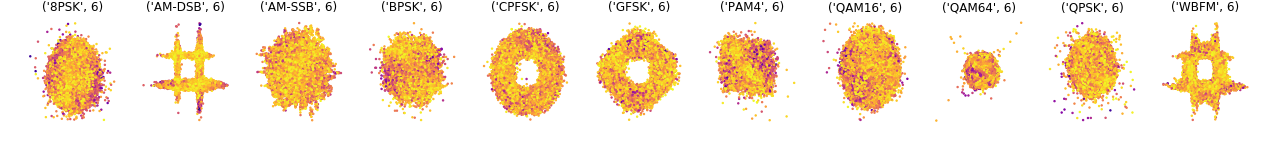} 
    \includegraphics[width=18.0cm,height=1.8cm]{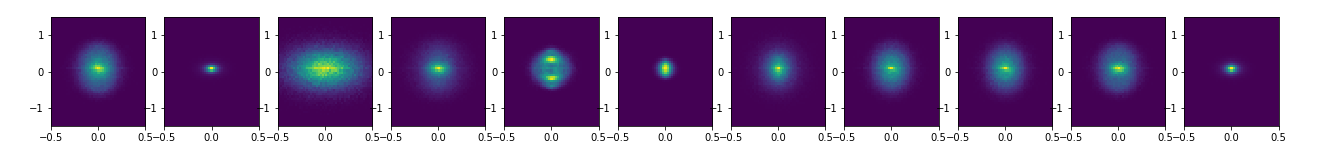} 
    \includegraphics[width=18.0cm,height=1.8cm]{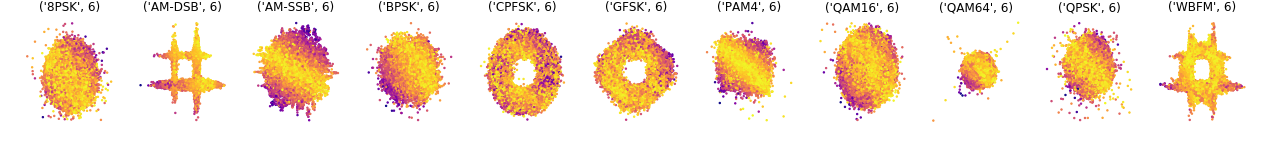} 
    \includegraphics[width=18.0cm,height=1.8cm]{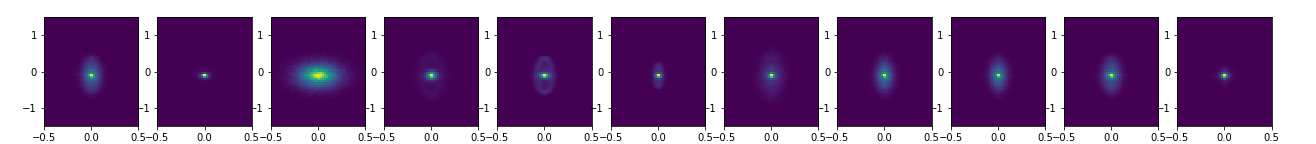} 
    \includegraphics[width=18.0cm,height=1.8cm]{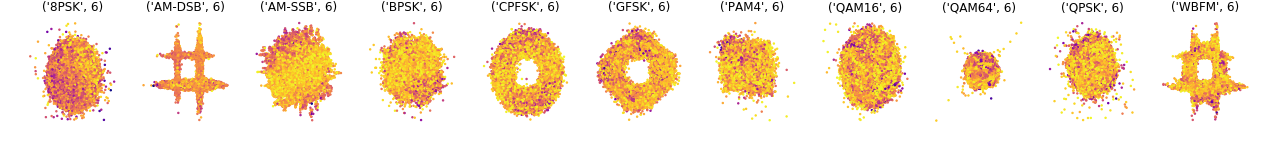} 
    \includegraphics[width=18.0cm,height=1.8cm]{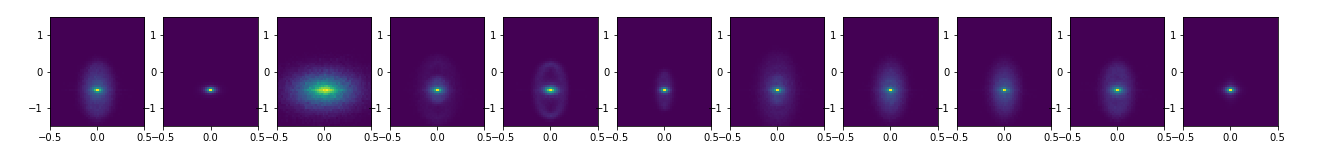} 
    \caption{Rows from top to bottom show the effect of adding more lags in the model, including lag difference with(out) correlation as features, removing one modulation from the training to see if network can predict it and the last row is the impact of SNR mismatch.}
\end{figure*}

\begin{figure*}
  \centering
    \includegraphics[width=5.0cm,height=3.0cm]{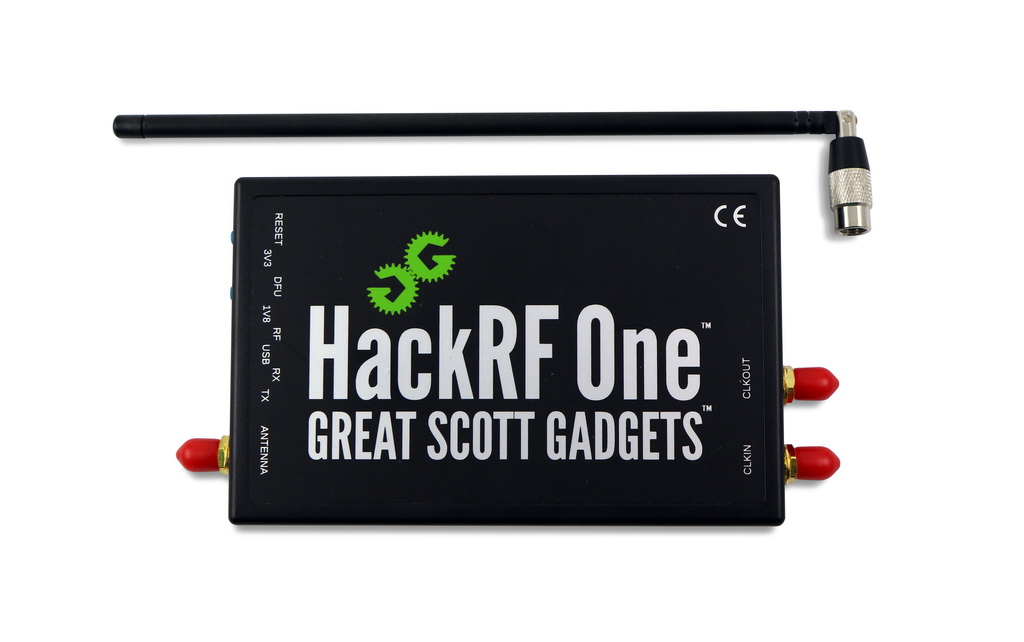}
    \includegraphics[width=10.0cm,height=4.0cm]{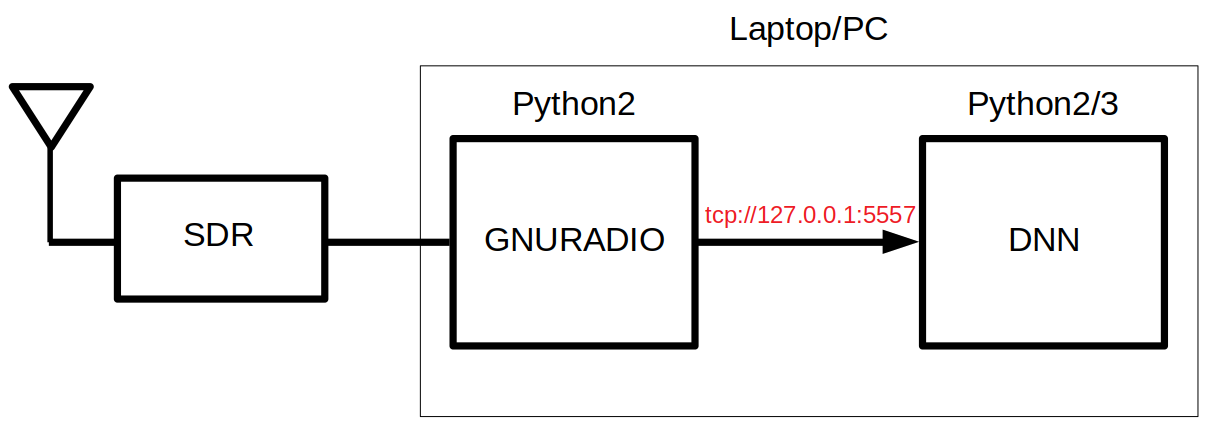}
    \caption{Block diagram of the experimental setup.}
\end{figure*}

\section{Conclusion}
Modulation recognition is addressed as an unsupervised learning problem here although it is shown in the literature that the performance of DNN based supervised learning is very good for classification. The embedded data later can be used for classification, clustering or detection. Although the autoencoder architecture is a very popular network for this task but for modulation recognition, it should be investigated what features and loss function can be added to the network in order to use the whole capacity of the architecture. Lag differences and correlations are introduced here for capturing the signal trajectory for a better embedding. Better embedding means similar signals end up close to and dissimilar far from each others. The loss function is quite easy to modify as it is done here based on Markov modelling. The effect of changing the lag and including extra features are discussed. Moreover, the effect of mismatch assuming two scenarios is studied. If a modulation and SNR were not part of training data, the network is robust enough to these cases and predict the unseen data. But embedded space is affected any how by all of them. The pipeline is evaluated by a cheap experimental setup at end.


%

\appendices
\section{Loss Function}
The loss function is defined as Eq. 2 in \cite{IEEEhowto:Hernandez}  

\begin{equation}
    LOSS=||x_{t+lag}-x^{'}_{t+lag}|| + ...
\end{equation}

However, any type of new term can be easily added to the loss function to see the effect of it in the training.


\section*{Acknowledgment}

This project is not supported by any official financial resources yet.

\ifCLASSOPTIONcaptionsoff
  \newpage
\fi



%

%

\begin{IEEEbiography}{Amin Abbasloo}
is a computer science engineer working on deep learning, machine vision and software development.
\end{IEEEbiography}
\begin{IEEEbiography}{Alan Salari}
is an electrical engineer working on RF techniques and hardware development.
\end{IEEEbiography}





\end{document}